\documentclass[11pt,twoside]{article}
\usepackage{fancyhdr}
\usepackage[colorlinks,citecolor=blue,urlcolor=blue,linkcolor=blue,bookmarks=false]{hyperref}
\usepackage{amsfonts,epsfig,graphicx}
\usepackage{afterpage}
\usepackage{comment}
\usepackage{amsmath,amssymb,amsthm} 
\usepackage{fullpage}
\usepackage[T1]{fontenc} 
\usepackage{epsf} 
\usepackage{graphics} 
\usepackage{amsfonts,amsmath}
\usepackage[sort]{natbib} 
\usepackage{psfrag,xspace}
\usepackage{color,etoolbox}

\newcommand{\var}{\text{var}}

\newcommand{\Pb}{\mathbb{P}}
\newcommand{\Pn}{\mathbb{P}_n}

\newcommand{\E}{\mathbb{E}}

\DeclareSymbolFont{bbold}{U}{bbold}{m}{n}
\DeclareSymbolFontAlphabet{\mathbbold}{bbold}

\theoremstyle{definition}

\theoremstyle{remark}

\newtheorem{remark}{Remark}

\begin{document}

\def\spacingset#1{\renewcommand{\baselinestretch}%
{#1}\small\normalsize} \spacingset{1}

\raggedbottom
\allowdisplaybreaks[1]

%%%%%%%%%%%%%%%%%%%%%%%%%%%%%%%%%%%%%%%%%%%

  \title{\vspace*{-.3in} {Discussion of ``On nearly assumption-free tests of  nominal confidence interval coverage for causal parameters  estimated by machine learning''  by Lin Liu, Rajarshi Mukherjee, and James Robins }}
  \author{\\ Edward H.\ Kennedy, Sivaraman Balakrishnan, Larry A.\ Wasserman \\ \\
    Department of Statistics \& Data Science \\
    Carnegie Mellon University \\ \\ 
    \texttt{\{edward, siva, larry\} @ stat.cmu.edu} \\
\date{}
    }

  \maketitle
  \thispagestyle{empty}

\vspace{.1in}

\noindent

\section{Introduction}

We congratulate the authors on their exciting paper, which introduces a novel idea for assessing the estimation bias in causal estimates. Doubly robust estimators are now part of the standard set of tools in causal inference, but a typical analysis stops with an estimate and a confidence interval. The authors give an approach for a unique type of model-checking that allows the user to check whether the bias is sufficiently small with respect to the standard error, which is generally required for confidence intervals to be reliable. \\

We begin our comments
by looking at an example
of a simple functional.

\section{Expected Density Example} \label{sec:expdens}

In this section we illustrate the main ideas in the paper by applying them to a simpler functional. 
This allows us to understand better some of the critical insights of \citet{liu2020nearly}.
In particular we consider the classic expected density functional 
$$ \psi = \E\{p(X)\} = \int p(x)^2 \ dx . $$
This functional has been studied extensively, with estimation and inference by now well understood \citep{bickel1988estimating, birge1995estimation}. Further, although it is simple, it has many of the nice properties of more complicated functionals like the expected conditional covariance or average treatment effect. \\

An analog of a doubly robust estimator of $\psi$ is the one-step or first-order corrected estimator given by
$$ \widehat\psi = \frac{2}{n} \sum_{i=1}^n \widehat{p}(X_i) - \int \widehat{p}(x)^2 \ dx $$
where $\widehat{p}$ is an initial pilot estimator of the density $p$, which for simplicity is based on an independent auxiliary sample of size $n$. The rest of this
analysis is conditioned on this auxiliary sample. \\

To fix ideas we briefly summarize some results regarding the estimation of $\psi$ and the estimator $\widehat{\psi}$.
A simple calculation shows that we may write
\begin{align*}
\psi = \widehat{\psi} - 2 \left( \frac{1}{n} \sum_{i=1}^n \widehat{p}(X_i) - \mathbb{E}[\widehat{p}] \right) + \int (\widehat{p} - p)^2. 
\end{align*}
In rough terms, if $\int (\widehat{p} - p)^2$ is $o_p(1/\sqrt{n})$ then the first-order estimator achieves parametric rates (and is semi-parametrically efficient). As an example, over classical Sobolev or H\"{o}lder smoothness classes,
the first-order estimator is efficient if $s > d/2$, where $s$ denotes the smoothness parameter, and $d$ the dimension of the data.
On the other hand, it is well-known that a second-order U-statistic estimator \citep{laurent1996efficient} is semi-parametrically efficient if $s > d/4$ and otherwise achieves the minimax rate of $n^{-4s/(4s+d)}$.\\

To understand the work of \citet{liu2020nearly}, suppose we write our initial estimate as:
$$ \widehat{p} = \sum_{j=1}^\infty \widehat\theta_j \phi_j, $$
where the $\phi_j$ form an orthonormal basis with respect to the Lebesgue measure. Then for $p = \sum_j \theta_j \phi_j$ a straightforward calculation shows that the conditional bias (given the auxiliary sample) is
$$ \text{Bias} = \E(\widehat\psi - \psi) = - \int \Big\{\widehat{p}(x) - p(x) \Big\}^2 \ dx = - \sum_{i=1}^\infty (\widehat\theta_j-\theta_j)^2 . $$
Note that this is the bias, not the squared bias; therefore for this functional, the standard first-order estimator has the monotone bias property. 
We can decompose this bias as
$$ -\text{Bias} = \underbrace{\sum_{j=1}^k (\widehat\theta_j - \theta_j)^2}_{\text{Bias}_k} + \underbrace{\sum_{j>k} (\widehat\theta_j - \theta_j)^2}_{\text{Truncation Bias}} . $$
Then, in order to test if the estimator $\widehat\psi$ is biased, one can ignore the truncation bias and simply test if the first projected bias term is large. Of course, the projected bias is a quadratic function, and so can be estimated with a second-order estimator (in this case a simple U-statistic). For example, one could use a modified version of the estimator proposed by \citet{laurent1996efficient}:
$$ \widehat{\text{Bias}_k} = \sum_{j=1}^k \widehat\theta_j^2 - \frac{2}{n} \sum_{j=1}^k \widehat\theta_j \left[ \sum_{i=1}^n \phi_j(X_i) \right] + \frac{1}{n(n-1)} \sum_{j=1}^k \sum_{u=1}^n \sum_{v=1}^n \phi_j(X_u) \phi_j(X_v).$$
Using standard properties of U-statistics, or following the analysis of \citet{laurent1996efficient} we find that if $k = o(n)$ then $\widehat{\text{Bias}_k}$ is a $\sqrt{n}$-consistent estimator of $\text{Bias}_k$, and the standard 
Wald interval can be used to test hypotheses regarding the magnitude of $\text{Bias}_k$. \\

To summarize, in this very simple example we are able to see two key insights of \citet{liu2020nearly}:
\begin{enumerate}
\item For a wide class of functionals, the bias of the first-order estimator has a relatively simple form. Often a second-order U-statistic based estimator can be designed to estimate the bias (or a portion of the bias).
\item The statistician has considerable freedom in the setting of testing hypotheses regarding the bias of an estimator. In particular, in the above setting the statistician has license to 
both select a choice of an orthonormal basis, as well as to ignore the resulting ``truncation bias'', and focus more narrowly on testing the magnitude of $\text{Bias}_k.$
While these choices might affect the power of the resulting test, they do not affect the validity (Type I error control) of the test.
We return to this point in Sections~\ref{sec:structure} and~\ref{sec:probe}.
\end{enumerate}
The work of \citet{liu2020nearly} presents an elegant extension of these ideas to a more complex class of functionals that arise in causal inference.

\section{The Monotone Bias Property}

In this section we simply illustrate how the monotone bias property is a property of not only the functional but also a given estimator. In particular, given an estimator/functional pair satisfying the monotone bias property, we can construct a similar first-order estimator that  does \emph{not} have the monotone bias property, using the ideas of \citet{newey2018cross}, and also discussed by \citet{liu2020nearly} in Supplementary Materials Section S.1.1. \\ 

For example, consider the first-order estimator of the expected density from Section \ref{sec:expdens}
$$ \widehat\psi = 2\Pn(\widehat{p}) - \int \widehat{p}^2 $$
which, as noted previously, has conditional bias $\E(\widehat\psi-\psi)= - \int (\widehat{p} - p)^2$ and so satisfies the monotone bias property. An alternative first-order estimator is given by
$$ \widehat\psi_2 = \Pn(\widehat{p}_1 + \widehat{p}_2) - \int \widehat{p}_1 \widehat{p}_2 $$
where $\widehat{p}_j$ are two different density estimators, built from separate independent samples (this can always be achieved by splitting the sample into thirds rather than halves). Importantly, this alternative estimator has conditional bias equal to 
$$ \E(\widehat\psi_2 - \psi) = - \int (\widehat{p}_1 - p) (\widehat{p}_2-p) $$
and so does \emph{not} satisfy the monotone bias property.  The estimator $\widehat\psi_2$ has some nice properties: it is doubly-robust in the sense that it is consistent if either $\widehat{p}_1$ or $\widehat{p}_2$ is; it is root-n consistent, asymptotically normal, and efficient under the same conditions as the classic first-order estimator $\widehat\psi$; and further it can be shown that $\widehat\psi_2$ is minimax optimal when $p$ is H\"{o}lder-smooth if $\widehat{p}$ is a particular undersmoothed estimator.  \\

The same trick can be used to de-monotonize the expected conditional covariance: instead of using the standard first-order estimator
$ \Pn\left\{ (A - \widehat\pi)^2 \right\} $ we would instead use
$$ \Pn\left\{ (A-\widehat\pi_1) (A - \widehat\pi_2) \right\} $$
which has essentially the same properties  as $\widehat\psi_2$ (except that minimax optimality with undersmoothing in the H\"{o}lder setup would only be achieved when the H\"{o}lder smoothness of $\pi$ is greater than $d/4$, unless the density of $X$ was known). \\

Of course with these variants one could still use the approach detailed in Section 4 of \citet{liu2020nearly}, e.g., testing a null that the Cauchy-Schwarz bound on the bias is small. However this bound would no longer be as relevant for justifying small bias, since one would only need the typically weaker condition that, e.g., for the expected density, 
$$ - \int (\widehat{p}_1 - p) (\widehat{p}_2-p) = o_\Pb(n^{-1/2}) .  $$
Note for example that the mean of the left-hand-side above is the squared bias, which can be smaller than the mean squared error with undersmoothing. 

\section{The Role of Covariate Structure}

In this section we raise some questions regarding the role covariate structure should play in estimation and inference for causal effects. \\

Covariate structure plays a relatively important role in \citet{liu2020nearly}, as well as in the general higher-order influence function literature \citep{robins2008higher, robins2009semiparametric, robins2017minimax}. For example, minimax lower bounds for the average treatment effect and expected conditional covariance are known in models where the covariate density is known  \citep{robins2009semiparametric}; these bounds can be attained even when the density is unknown, but this requires sufficient smoothness of the density, at least whenever the propensity scores and regression functions are smooth, but not so smooth to admit root-n rates \citep{robins2008higher, robins2017minimax}. The complete minimax story with unknown density therefore appears to be an open problem. Based on upper bounds given in \citet{robins2008higher},  one might expect the rate to interpolate between the classic functional estimation rate $n^{-4s/(4s+d)}$ (where $s$ is the average smoothness of the propensity score and regression functions), and the slower fixed design rate $n^{-2s/d}$, e.g., an analog of which appears in \citet{wang2008effect} in a related conditional variance estimation problem. The connection between non-randomness (e.g., fixed design) and non-smoothness in the covariate density would be interesting to explore further; at the non-smooth extreme the aforementioned interpolation would suggest that these two structures align. \\

An interesting question is how and whether randomness of the covariates, and/or smoothness of the covariate density, should be relied upon in practice. In many problems one might expect convenience samples to be the norm, rather than true random samples, in which case it could be argued that the covariates should be conditioned on in any inferential procedures. In this kind of setup, one might wonder if estimators that exploit randomness/smoothness or other structure of the covariates might be less robust when such structure is not present; ideally our estimators would be able to adapt to such structure when present, but would otherwise be unaffected in the absence of randomness/smoothness. \\

More specifically, it would be interesting to explore whether higher-order influence function-based estimators have a robustness to misspecification of density estimates, e.g., relative to a standard first-order estimator, which does not require modeling the density. Similarly, it would be useful to know how higher-order estimators and/or the bias test proposed by \citet{liu2020nearly} perform in a fixed design setup, insofar as this is the right context for conditional-on-covariate inference. Note however that there are at least four possible choices of setups, based on the design and target parameter; for example for (half) the average treatment effect $\E(Y^1)$ one might consider:
\begin{enumerate}
\item the  parameter $\int \mu_1(x) \ d\Pb(x)$ in a usual random design setup;
\item  the parameter $\int \mu_1(x) \ dx$ in a fixed design  setup with $X$ equally spaced on a grid;
\item[3/4.] the parameter $\frac{1}{n} \sum_i \mu_1(x_i)$ in either a random or fixed design setup.
\end{enumerate}

\bigskip

Since the fixed design setup leads to slower rates, e.g., in variance estimation, one might expect this to affect the power but not validity of the bias test of  \citet{liu2020nearly}. However the details are not obvious to us.

\section{Exact Parametric Inference}

The authors focus on nonparametric inference, and we agree that this is usually the best way to proceed.
Still, it is interesting to ask what happens in a parametric model.
Typically such models are too inflexible to yield reliable inference.
But suppose we use a highly flexible parametric family.
Examples include mixture models and exponential families based on truncated series expansions.
The extra flexibility these models provide come at a cost: constructing confidence intervals can be even harder
than in the nonparametric case. Indeed, models such as mixture models are non-regular and inference is challenging.
However, a recent method called {\em universal inference} \citep{wasserman2020universal}
provides a potential solution. The method is simple. Denote the model by $\mathcal{P} = \{p_\theta:\ \theta\in\Theta\}$.
Split the data  into two: ${\cal D}_0$ and ${\cal D}_1$.
Let 
\begin{align*}
T(\theta)= \frac{{\cal L}_0(\widehat\theta_1)}{{\cal L}_0(\theta)},
\end{align*}
where ${\cal L}_0$ is the likelihood constructed from ${\cal D}_0$ and $\widehat\theta_1$ is any estimate of $\theta$ based on ${\cal D}_1$.
This could be the maximum likelihood estimator, a robust estimator or a minimum distance estimator for example. 
Now repeat this process $B$ times yielding $T_1(\theta),\ldots, T_B(\theta)$ and define $\overline{T}=B^{-1}\sum_b T_b(\theta)$.
Finally, set $C_n = \{ \theta:\ \overline{T}(\theta)\leq 1/\alpha\}$. Then it was shown that 
\begin{align*}
P_{\theta^*}(\theta^* \in C_n)\geq 1-\alpha,
\end{align*} 
for all $\theta^* \in \Theta$. The result extends to functionals.
Let $\psi = f(\theta^*)$ and replace the term ${\cal L}_0(\theta)$ with the profile likelihood for $\psi$.
Then we get a finite sample valid confidence set for $\psi$.
It would be interesting to compare exact, flexible parametric methods to approximate asymptotic methods in terms of coverage, robustness, and efficiency.

\section{Structure-Driven vs.\ Methods-Driven Estimation \& Inference}
\label{sec:structure}
In this section we draw a distinction between structure-driven and methods-driven estimation and 
inference, and relate this to the work of \citet{liu2020nearly}  and \citet{robins2008higher, robins2009semiparametric, robins2017minimax}. \\

Many successful tools in statistics and machine learning are algorithmic or methods-driven,
i.e., were not derived with a clear inferential goal (for example optimality or minimaxity in a particular statistical model). The most successful of these 
tools are often ``robust'', in the sense of having strong guarantees under many different statistical models. Prominent examples include methods like spectral clustering, random forests, and deep neural networks. 
On the other hand, other methods are decidedly more model-based or structure-driven: for example, a truncated series estimator for regression or density estimation, where one assumes some structure (e.g., decay of coefficients in a particular basis, H\"{o}lder smoothness, etc.) and then tailors the estimator precisely to exploit this structure. These structure-driven estimators 
are often crucial for setting theoretical benchmarks, but their theoretical guarantees may not be as "robust". This is reminiscent of the discussion in \citet{donoho1995wavelet} (Section 3) distinguishing between estimators that are exactly minimax in a single model, versus approximately minimax across a large collection of models. The distinctions between structure and methods-driven methods are sometimes murky, but we find them nonetheless useful. One might attempt to similarly categorize classical first-order methods, as well as their higher-order counterparts  \citep{liu2020nearly, robins2008higher, robins2009semiparametric, robins2017minimax}, both in the context of functional estimation, as well as for the bias tests of \citet{liu2020nearly}. \\

In the context of estimation, standard first-order estimators, such as the classic doubly robust estimator of the average treatment effect, seem to us more methods-driven. They are a ``black-box'' correction that can be applied to any initial nuisance function estimates, and in general yield improved guarantees over the plug-in estimator.
This at least partially explains the strong empirical success these methods have found across various application domains. On the other hand, higher-order influence function methods typically rely on carefully constructed series estimates and achieve better performance, e.g., over appropriate H\"{o}lder spaces, potentially at the expense of being more structure-driven. Higher-order estimators are useful in understanding the fundamental limits for functional estimation, but may be less agnostic about the underlying structure than corresponding first-order estimators, and may require more significant domain-knowledge in their practical application. \\

The previous discussion was focused on first and higher-order methods for estimation. 
However, \citet{liu2020nearly} address the slightly different problem of testing. To simplify the discussion we focus on the simplest cases, where the monotone-bias property holds, and the claims of \citet{liu2020nearly} are the strongest.
As we alluded to earlier, the bias testing problems that \citet{liu2020nearly} consider have some interesting features: 
\begin{enumerate}
\item The statistician has flexibility to specify the (portion of the) bias she wishes to test for.
\item As is typical in hypothesis testing problems the primary focus is on controlling the Type I error (ensuring validity). In reasoning about the power of the test, the statistician also has the flexibility to specify the alternatives against which she desires power.
\item The authors work in a ``falsification'' setup, where in essence any rejection is useful from a practical standpoint in tempering the statistician's claims about the first-order estimator and its associated Wald interval, and any failure to reject comes with few added harms.
\end{enumerate}
From this standpoint, we can revisit the methods-driven versus structure-driven distinction, and ask: are there  downsides to using structure-driven methods for inference? We believe the answer may be yes, but more work is required. In the next section we consider a simple extension that could help move the \citet{liu2020nearly} approach to be more methods-driven.

\section{A Simple Ensembled Bias Test}
\label{sec:probe}

In this section we propose a simple modification of the bias test that may provide some practical advantages in terms of power, by incorporating generic flexible regression tools, and thus weakening ties to classical series methods. \\

In the usual series setup, one specifies a basis that is a priori expected to capture underlying structure well (typically H\"{o}lder smoothness), such as Daubechies wavelets or a Fourier series. We instead suggest an alternative based on trying to estimate a good basis from the data, by constructing a dictionary of appropriate predictors, and then orthogonalizing them. The motivation behind this is that we want the truncation bias (the second term in Equation 2.3) to be small in order for the test to have power, and this depends on the basis terms capturing enough  important structure in the projected bias.    \\

For simplicity consider the expected conditional variance functional $\E\{\var(A \mid X)\}$.  Suppose we have access to an auxiliary sample of observations (note we could always split the sample accordingly so as to  have one more fold).  In the auxiliary sample we regress the residuals $(A-\widehat{b}(X))$ on $X$, using $m$ different methods. For example we could use random forests, lasso, boosting, etc.\ as well as variants of these and other methods with different tuning parameter choices. Importantly, these methods could capture different kinds of structure, e.g., smoothness, sparsity, additivity, etc. This yields $m$ different predictors $(\widehat{f}_1,..., \widehat{f}_m)$. Next we evaluate these predictors on the sample used for testing the bias, to obtain $m$ different $n$-vectors of predicted values $\widehat{f}_j(X_i)$ for $j=1,...,m$ and $i=1,...,n$. Finally we orthogonalize (the span of) these vectors and use them as an estimated basis; namely using the notation in the paper we let $\overline{z}_k(X)=(\widehat{f}_1(X),...,\widehat{f}_m(X))$, possibly concatenated with a typical fixed basis. \\

Intuitively since the $m$ different predictors can potentially capture diverse kinds of structure, if one (or a weighted combination) of predictors leads to small truncation bias then the projection onto their span should as well \citep{tsybakov2003optimal}.  It seems that the theory from  \citet{liu2020nearly} would all go through as before, since conditional on the auxiliary sample the estimated basis functions are all fixed; the advantage is there should be some gain from tailoring the basis functions to the specific bias structure that is trying to be captured. \\

\begin{remark}

Lastly,
we would like to 
pick a small nit.
The authors have taken the now common approach
of referring to statistical prediction methods
as ``machine learning.''
Perhaps the field of statistics should instead re-claim regression.
Further, the field of machine learning is large and diverse, with many successes outside of pure prediction problems; so equating the term machine learning with prediction tools diminishes these other successes. \\

\end{remark}

\section*{References}
\vspace{-1cm}
\bibliographystyle{abbrvnat}
\bibliography{/Volumes/flashdrive/research/bibliography}

\end{document}